\documentstyle[12pt,epsfig]{procsla}

\newcommand{\beq}{\begin{eqnarray}}
\newcommand{\eeq}{\end{eqnarray}}
\newcommand{\no}{\nonumber}

\newcommand{\vp}{v\!\!\:\cdot\!\!\: p}

\newcommand{\vD}{v\!\!\:\cdot\!\!\:{D}}

\def\slash#1{#1 \hskip -0.5em / }

\def\Pp{\frac{1 + \slash{v}}{2}}

\def\gl#1{Eq.~(\ref{#1})}

\def\L#1{{\cal L}_{\mbox{\scriptsize #1}}}

\def\Tr#1{{\rm Tr}\left[#1\right]}

\begin{document}
\begin{flushright}
hep-ph/9606451
\end{flushright}
\vspace{2em}

\begin{center}
{\bf RECENT RESULTS IN THE NJL MODEL WITH HEAVY QUARKS\footnote
{Talk presented at the III.\ German-Russian Workshop on
 {\it Theoretical Progress in \sc Heavy Quark Physics},
  Dubna, 20-22 May 1996. To appear in the conference proceedings.}}

\vspace*{1cm}
THORSTEN FELDMANN\footnote{Supported by
{\it Deutsche Forschungsgemeinschaft} under contract Eb 139/1--2.}
\\
{\it Institute of Physics, Humboldt University,
Invalidenstra\ss{}e~110,\\ 10115~Berlin, Germany\\}
\end{center}

\vspace*{1.5cm}
\begin{abstracts}
{\small We investigate the interplay of chiral and heavy quark
symmetries by using the NJL quark model.
Heavy quarks with finite masses $m_Q$ as
well as the limit $m_Q \to \infty$ are studied. We found 
large corrections to the heavy mass scaling law for
the 
pseudoscalar decay constant.
The influence of external momenta
on
the shape parameters of the Isgur-Wise form factor is discussed.}
\vspace*{1.5cm}
\end{abstracts}

\section{Introduction}


 
As is well known 
for heavy quark masses
$m_Q \gg \Lambda_{QCD}$ one can expand the QCD lagrangian
connected with heavy quark dynamics in terms of $1/m_Q$,
leading to
\beq
\L{HQL} &=&
\bar Q_v \, (i \vD ) \, Q_v
+  {\cal K}_v + {\cal M}_v + O(1/m_Q^2) \ ,
\eeq
with $D_\mu$ being the QCD covariant derivative
and ${\mathcal K}_v$ and ${\mathcal M}_v$
being the kinetic and chromomagnetic energy
of the heavy quark, respectively\cite{MRR}.
In the heavy quark limit $m_Q \to \infty$ (HQL)
 the contribution of the latter vanishes, and the remaining
lagrangian is independent of the heavy quark flavor and
spin.
Consequently, 
heavy mesons are organized in spin symmetry doublets
with $J= j_l \pm 1/2$ where $j_l$ is the spin
of the light degrees of freedom\cite{Falk}.
%

In the sector of light quark flavors
$q=(u,d,s)$, QCD possesses an approximate $SU(3)_L \times
SU(3)_R$ chiral symmetry which is spontaneously broken
to $SU(3)_V$, leading to the emergence of (pseu\-do)\-Gold\-stone
bosons $\pi,K,\eta$.
We use
the common non-linear representation, where the Goldstone bosons
(denoted by $\pi$) and their transformations under
chiral symmetry $SU(3)_L \times SU(3)_R$ are given by
\beq
&&
 \xi = e^{i\pi/F} \to L \, \xi \, U^\dagger = U \, \xi \, R^\dagger \ ,
\eeq
which defines the matrix $U(\pi,L,R)$ as a non-linear function of
its arguments.
Then heavy meson fields $\Phi$ transform
under chiral symmetry as\cite{DonoghueWise,Casalbuoni}
$\Phi \to \Phi \, U^\dagger $.

\section{The NJL model with heavy quarks}

We will use the above symmetry considerations
in order to construct the NJL quark model as a low-energy
approximation to QCD.
One choses suitable 4-quark operators
 representing the scalar, pseudoscalar,
vector and axial-vector channel, respectively,
\beq
\L{}^{\bar q Q} &=&
 2 \, G_3 \, \left( 
 	(\bar Q \, q) \, (\bar q \, Q)
	+ ( \bar Q \, i \gamma_5 \, q )\, (\bar q \, i \gamma_5 \, Q)
  \right. \no 
\\
&& \qquad \left. 
	- \frac{1}{2} \, (\bar Q \, \gamma_\mu \, q) \,
			(\bar q \, \gamma^\mu \, Q)
        - \frac{1}{2} \, (\bar Q \, i \gamma_\mu\gamma_5 \, q) \,
			(\bar q \, i\gamma_5\gamma^\mu \, Q)	
       \right) 
\no \\
\stackrel{m_Q \to \infty} \longrightarrow
\L{}^{\bar q Q_v} &=&
G_3 \, \left( ( \bar Q_v \, i \gamma_5 \, q )\, (\bar q \, i \gamma_5 \, Q_v)
 	     - (\bar Q_v \, \gamma_\mu^\perp \, q) \,
			(\bar q \, \gamma^\mu_\perp \, Q_v)
	  \right. \no 
\\
&& \qquad \left. 
	+ ( \bar Q_v \, q )\, (\bar q \, Q_v)
 	     - (\bar Q_v \, i\gamma_\mu^\perp\gamma_5 \, q) \,
			(\bar q \, i\gamma_5\gamma^\mu_\perp \, Q_v)
       \right) 
\ .
\label{LqQ}
\eeq
Here the coupling constant $G_3$ has dimension~(-2) and is
related to an UV cut-off $\Lambda$, reflecting
the scale of chiral symmetry breaking of the order of 1~GeV.
Note that 
it is 
the heavy quark residual
momentum $(P_Q - m_Q \, v^\mu)^2 \sim O(p_q^2) < \Lambda^2$
which is regularized, and in this way
the NJL model may indeed be applied.
Note also, that all reference to gluons is gone, and 
consequently the model has no idea about confinement.

The above lagrangian can be transformed {\em exactly}\/
into a quark-meson interaction by integrating in
auxiliary fields, which in the HQL leads to 
\beq
\tilde \L{}^{\bar \chi Q} 
& \stackrel{m_Q \to \infty} \longrightarrow &
- \bar Q_v \, \left[ H + K \right] \, \chi
- \bar \chi \, \left[ \bar H + \bar K \right] \, Q_v
+ \frac{1}{2 G_3} \,
\Tr{\bar H  H - \bar K K}
\label{eq5}
\ .
\eeq
Here we introduced the above mentioned non-linear representation
by rotating the light quark fields
$\chi_R = \xi \, q_R $ , \ 
	$\chi_L = \xi^\dagger \, q_L $.
The Dirac matrices for the spin-symmetry doublets with
$J^P = (0^-,1^-)$ and $J^P = (0^+,1^+)$, respectively,
are given by
\beq
 H &\equiv& 
\Pp 
\, ( i \Phi_5 \gamma_5 + \slash\Phi ) \ , \qquad
 K \equiv 
\Pp 
\, ( \Phi + i \slash\Phi'_{5}\gamma_5 )  \ ,
\eeq
with $v_\mu \, \Phi^\mu = v_\mu \, \Phi_5^\mu = 0$.
The quark fields are now integrated out explicitly by a functional Gaussian 
integration. The real part of the
resulting quark determinant 
after Wick-rotation ($ D \to D_E$)
is regularized by
a proper-time integral 
\beq
\ln \left| \det D \right|
&\to&
- \frac{N_c}{2} \, \int_{1/\Lambda^2}^{1/\mu^2} \,
\frac{ds}{s} \, \int \Tr{ e^{-s \, D_E^\dagger  D_E}}
\ .
\label{reg}
\eeq
Here, $\mu$ serves as an adjustable parameter and may be
interpreted as the scale 
up to which the quarks have been integrated out.
Especially, at $\mu = \Lambda$ the contribution of the 
quark determinant vanishes by construction, and we recover
the original interaction of quarks with static meson fields
in \gl{eq5}. 

It is a straight forward task to generalize the presented
ideas to heavy mesons of higher excitations like the
$J^P=(1^+,2^+)$ multiplet.
It can be represented as\cite{ChoKoerner}
\beq
T^\mu &=& \frac{1 + \slash{v}}{2} \,
      \left\{ {\Phi}^{\mu\nu} \gamma_\nu +
      \sqrt{\frac{3}{2}} {\Phi_5}^\nu \, i\gamma_5\,
   \left(g_{\mu\nu} - \frac{1}{3} \gamma_\nu (\gamma_\mu -
v_\mu)\right)
  \right\}
\eeq
with $v_\nu \Phi^{\mu\nu} = v_\nu {\Phi_5}^\nu = 0$,
$\Phi^{\mu\nu}= \Phi^{\nu\mu}$ and
$\Phi^\mu_\mu = 0 $.
For the light degrees of freedom with $j_l = 3/2$ we
construct a Rarita-Schwinger representation
 from the light quark fields, introducing a covariant
derivative 
$i {\cal D}_\nu \equiv i \partial_\nu + i/2 \,
(\xi^\dagger \partial_\nu \xi + \xi \partial_\nu \xi^\dagger)$
of the chiral $SU(3)_V$, 
\beq
\tilde \L{}^{\chi Q_v} &\to&
- \bar Q_v \, 
T^\nu \, i \!\stackrel{\rightarrow} {\cal D}_\nu  \chi
+ \bar \chi \, 
 i \!\stackrel{\leftarrow} {\cal D}_\mu
 \bar T^\mu \, Q_v
- \frac{1}{2 G_4} \, \Tr{\bar T^\mu T_\mu}
\ ,
\eeq
introducing a new coupling constant $G_4 \sim O$(1~GeV)$^{-4}$.
For further details we refer to our original
works\cite{we,next}.

\section{Results}
\begin{figure}[htb]
\begin{center}
\psfig{file = 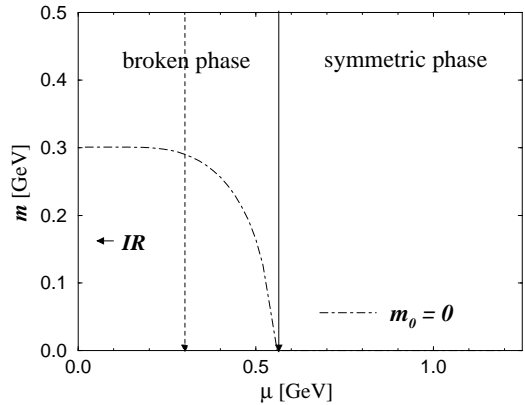, bb = 50 70 610 720, width=6cm, angle = -90}
\end{center}
\caption{\small The solutions  $m_q$ of \gl{gap} as
an implicit function of the scale $\mu$.} 
\label{gapfig}
\end{figure}

For low $\mu$ the contribution of the quark determinant 
leads to dynamical breaking of chiral symmetry. In the
NJL model this 
is governed by the gap-equation for the light quark
constituent mass\cite{EbRe}
\beq
m_q = m_q^{(0)} + 8 \, G_1 \, m_q \, I_1
\ , \qquad
I_1 = \frac{N_c}{16\,\pi^2} \,\left(
		\Lambda^2 - \mu^2 + O(m_q^2) \right) 
\label{gap}\ .
\eeq
Fig.~\ref{gapfig} shows the behaviour of the solution
for $m_q$ as a function of the scale $\mu$
in the chiral limit, i.e.\ for a vanishing current quark mass
$m_q^{(0)} = 0$.
We may use $\mu$ for separating the IR region of 
the proper-time integral in \gl{reg} -- which
is believed to be governed by confinement and can not be
well described in our model -- from the region where 
the model is assumed to lead to reasonable
results, namely to a consistent description
of dynamical chiral symmetry breaking.
The dashed vertical line indicates the convenient
choice $\mu = 300$~MeV.
Note that in the chiral limit the critical scale for
the phase transition can be read off analytically from
\gl{gap}
\beq
\mu_c^2 &=& \Lambda^2 - \frac{2 \, \pi^2}{3 \, G_1}
\approx (550~\mbox{MeV})^2 \ , \qquad 
(G_1 = 5.25~\mbox{GeV}^{-2} , \
 \Lambda = 1.25~\mbox{GeV})\mbox{\cite{we}}
\ .
\eeq


With the self-energy contributions of the quark-determinant 
for scales $\mu < \Lambda$ the several meson fields in the
NJL model become dynamical degrees of 
freedom. 
We stress, that the limit $\mu \to 0$ in the proper-time
integrals is in general 
not suitable if the external momenta exceed
the sum of the internal quark masses since by analytic
continuation the self-energy then receives an imaginary part,
which by the optical theorem is
connected with unphysical quark-antiquark thresholds.

The mass-spectrum of the heavy mesons in the NJL model is
then determined by the value of the 4-quark coupling constant
$G_3$ (with all other parameters fixed in the light meson
sector). 
The NJL model further gives 
simple predictions for the weak decay constants of the
heavy pseudoscalar ($P$) and vector
mesons ($V$) in terms of the wave function renormalization
factors\footnote{Note that in the HQL a
factor of $M_H$ is absorbed into the normalization
of the heavy meson fields.} $Z$ and the coupling constant $G_3$,
\beq
\left.
\begin{array}{l}
\sqrt{M_P} \, f_P = (1+ \delta) \, 
 \frac{\sqrt{Z_P/M_P}}{G_3}
\\
\sqrt{M_{V}} \, f_{V} = \frac{\sqrt{Z_{V}/M_{V}}}{G_3}
\end{array}   
\right\} 
&\stackrel{\rm HQL} \longrightarrow 
& \sqrt{M_H} \, f_H = \frac{ \sqrt{Z_H} }{ G_3}
\label{fH}
\eeq
Here, $\delta$ denotes the contribution from the
mixing of the pseudoscalar and axial-vector states
beyond the HQL.
The perturbative corrections to \gl{fH} are known;
in the leading-log approximation one has\cite{LLog}
$f_D = f_B \, \left(\alpha_s(m_c)/\alpha_s(m_b)\right)^{6/25}$.

The results are collected in Table~\ref{table}.
Compared to experiment 
the $SU(3)_F$-mass splitting
due to the different light quark masses $m_u = 300$~MeV
and $m_s = 510$~MeV is somewhat underestimated. 
The experimental fine-splitting between the spin-symmetry
partners is recovered for rather small values of $m_b$
and $m_c$. We further observe large
corrections to the HQL scaling law
for the pseudoscalar decay constant,
whereas the deviations for the vector decay constant
are within the usual expectation. This is mostly
due to the large value of the mixing contribution
$\delta$ which obtains values up to 50\% in
the case of $D$-mesons.  
The values for the weak decay constants are on the
lower side of the range obtained from lattice results\cite{wittig}.

\begin{table}[bth]
\caption{\small Heavy Meson masses and decay constants 
	in the NJL model for a) the HQL and b) finite values
	of $m_Q$. Here $\bar\Lambda$ is the mass-difference
	between heavy meson and heavy quark in the HQL. The
	heavy quark masses are fitted to the averaged heavy
	meson masses $\bar M_H = 1/4 \, (3 \, M_{V} + M_P)$.
	These are then used to estimate the mass-splittings
	$\delta_H = M_{V} - M_P$ and the decay constants.}
\label{table}
\begin{center}

a)
\begin{tabular}[t]{l|ccc|cc}
\hline
$G_3$ 
& $\bar\Lambda_u$ & $\bar\Lambda_s - \bar\Lambda_u$ & $M_K - M_H$
& $f_B$ & $f_{B_s}/f_B$ 
\\
{}[GeV]$^{-2}$ & \multicolumn{3}{c|}{[MeV]} & [MeV] & \\
\hline \hline
6.3  & 300 & 60 & 245 & 140 & 1.04 \\ 
4.5  & 400 & 75 & 190 & 150 & 1.04 \\ 
2.9  & 500 & 80 & 140 & 160 & 1.03 
\\
\hline
Exp.\cite{PD} & & 100 &  &  &
\\ \hline
\end{tabular}\\
\vspace{1em}
b)
\begin{tabular}[t]{l|cccc|cccc}
\hline 
$G_3$ 
& $m_b$ & $m_c$ & $\delta_B$ & $\delta_D$ 
& $f_B$ & $f_{B^*}$ & $f_D$ & $f_{D^*}$ 
\\
{}[GeV]$^{-2}$ & \multicolumn{2}{c}{[GeV]} &
\multicolumn{2}{c|}{[MeV]} &
\multicolumn{4}{c}{[MeV]} \\
\hline \hline
6.3 & 4.97 & 1.53 & 81 & 239 & 130 & 135 & 160 & 200  \\
4.5 & 4.88 & 1.43 & 63 & 192 & 140 & 145 & 170 & 225  \\
2.9 & 4.78 & 1.33 & 46 & 146 & 150 & 155 & 190 & 260
\\
\hline 
Exp.\cite{PD} & & & 46 & 141 & 
\\ \hline
\end{tabular}
\end{center}
\end{table}


The NJL result for the Isgur-Wise function is obtained
from the quark determinant by 
insertion of a heavy quark current.
We present the NJL results for 
the slope parameter $\rho = \sqrt{-\xi'(1)}$ and the
curvature $c_0 = \xi''(1)/2$,
\beq
\vp = 300~\mbox{MeV} :& \quad \rho = 0.84 \ , \ \ & c_0 = 0.49 \no \ ; \\
\vp = 400~\mbox{MeV} :& \quad \rho = 0.94 \ , \ \ & c_0 = 0.72 \no \ ; \\
\vp = 500~\mbox{MeV} :& \quad \rho = 1.07 \ , \ \ & c_0 = 1.18
\ .
\eeq
Note that here the values of $\vp = \bar\Lambda$ and $\rho$
and $c_0$
are positively correlated. This should be compared
with the estimated theoretical relation\cite{neubert} 
$c_0 \simeq 0.72 \, \rho^2 - 0.09$ together
with the phenomenological bounds 
$0.84 \leq \rho \leq 1.00$.

\section{Conclusions}

We have presented the 
NJL quark model as a low-energy approximation to
QCD including the approximate symmetries for light
and heavy flavors. 
It has been shown that this model is equivalent to a relativistic
quark-meson model for the several heavy meson states.
We have introduced an adjustable scale parameter $\mu$
in the proper-time regularized quark determinant, such that
at $\mu=\Lambda$ the model matches to
the case of static meson fields which become
then dynamical for $\mu < \Lambda$. 
The dynamical
breaking of chiral symmetry occurs around $\mu \approx 550$~MeV. 
We then chose
a finite value of $\mu \approx 300$~MeV in order to separate
the IR region of the proper-time integration.
With this one obtains reasonable results for the
heavy meson mass spectrum both, with finite and infinite heavy
quark masses, for the decay constants and for the Isgur-Wise
form factors. 

\vspace{2em}

\noindent{\bf Acknowledgements}
\vspace{1em}

{\it 
I would like to express my gratitude to the Organizing Committee
of the workshop and the people at the JINR Dubna for their
warm hospitality as well as to the Heisenberg-Landau Program
for financial support.
It is also a pleasure to thank D.~Ebert,
H.~Reinhardt and R.~Friedrich for their
contributions to
the results presented here. 
}


\end{document}